\documentclass[twocolumn,aps,prb]{revtex4}
\usepackage{bm}
\usepackage{graphicx}
\usepackage{epsfig}
\begin{document}
\title{Theory of dual fermion superconductivity in hole-doped cuprates}
\author{Jun Chang$^{1,2}$, Jize Zhao$^3$ }
\affiliation{$^{1}$ College of Physics and Information Technology, Shaanxi Normal University, Xi'an 710119, China,$^{2}$Institute of Theoretical Physics, CAS, Beijing 100190, China,\\ $^{3}$Institute of Applied Physics and Computational Mathematics, Beijing 100088, China. $^*$e-mail: jun.chang@hotmail.com}
\date{\today}

\begin{abstract}
Since the discovery of the cuprate high-temperature superconductivity in 1986, a universal phase diagram has been constructed experimentally and numerous theoretical models have been proposed. However, there remains no consensus on the underlying physics thus far. Here, we theoretically investigate the phase diagram of hole-doped cuprates based on an itinerant-localized dual fermion model, with the charge carriers doped on the oxygen sites and localized holes on the copper $d_{x^{2}-y^{2}}$ orbitals. We analytically demonstrate that the puzzling anomalous normal state or the strange metal could simply stem from a free Fermi gas of carriers bathing in copper antiferromagnetic spin fluctuations. The short-range high-energy spin excitations also act as the `magnetic glue' of carrier Cooper pairs and induce $d$-wave superconductivity from the underdoped to overdoped regime, distinctly diffrent from the conventional low-frequency magnetic fluctuation mechanism. We further sketch out the characteristic dome-shaped critical temperature $T_c$ versus doping level. The emergence of the pseudogap is ascribed to the localization of partial carriers coupled to the local copper moments or a crossover from the strange metal to a nodal Kondo-like insulator. Our work provides a consistent theoretical framework to understand the typical phase diagram of hole-doped cuprates and paves a distinct way to the studies of both non-Fermi liquid and unconventional superconductivity in strongly correlated systems.
\end{abstract}

\maketitle

\section{\label{intro}Introduction}

The Bardeen-Cooper-Schrieffer (BCS) theory reveals the phase transition from a Landau Fermi liquid into superconductivity in conventional superconductors \cite{Bardeen1957}.  In this theory, the electron quasiparticles around the Fermi surface are bound into coherent Cooper pairs in the presence of an attractive potential resulted from electron-lattice interaction. However, for unconventional superconductors, such as cuprates \cite{Bednorz1986}, iron-pnictides \cite{Kamihara2008} and heavy fermions \cite{Steglich1979}, the normal states are often non-Fermi liquids \cite{Gurvitch1987,Thomas1988,Seaman1991,Dagotto1994,Stewart2001,Stewart2006,Lee2006,Scalapino2012,Keimer2015}, which is seemingly indescribable in terms of weakly interacting quasiparticles. In addition, phonons are unlikely to mediate electrons into Cooper pairs in these strongly correlated systems \cite{Lee2006,Pfleiderer2009,Scalapino2012}. To drive the phase transition, the sought bosons or bosonic excitations should already exist in the normal state by analogy with phonons in Fermi-liquid metals. The spin fluctuation is believed to be the most promising candidates \cite{Dahm2009,Carbotte2011,Scalapino2012,Scalapino1986,Monthoux1991,Monthoux1992,Moriya1990,Millis1990,Chubukov2008}. However,  so far, the most serious challenge to the theories is to elucidate the non-Fermi liquids and superconductivity on the same footing.

In these unconventional superconductors, one striking feature is the coexistence of localized and itinerant electrons \cite{Park2008}. For example, in heavy fermion compound UPd$_2$Al$_3$, the dual $f$ electron model successfully explains both the superconductivity and the magnetic resonance based on the interaction between the localized and itinerant $f$ electrons \cite{McHale2004,Chang2007}. Likewise, the duality of electrons is shared by hundreds of cuprate superconductors besides a universal phase diagram and a layered structure made up of one or more copper-oxygen (Cu-O) planes \cite{Rybicki2016,Haase2009,Muller2007,Walstedt1994,Curro1997,Li2012Two,Barzykin2009}. In the insulating parent compound of cuprates, only the $d_{x^2-y^2}$ orbital is half-filled in each Cu ion, denoted by a localized hole with spin-$1/2$. With the hole doping, primarily on O sites, the introduced carriers gradually melt antiferromagnetic order and eventually lead to superconductivity. Generally, it is believed that the physical properties are dominated by the Cu-O planes due to the universality in cuprates. Thus, a three-band model is constructed to include both localized and conducting electrons \cite{Emery1987}. It can be further reduced to a simpler one, such as the Kondo-Heisenberg model \cite{Zaanen1988}, the spin-fermion \cite{Chubukov2008} and $t$-$J$ model \cite{Zhang1988}, which are widely used to describe the itinerant-localized duality of electrons in cuprates.

\section{Dual fermion model }
Here, we take advantage of a dual fermion model or Kondo-Heisenberg-like model to explore the generic cuprate phase diagram. Markedly different from the one-component spin-fluctuation models proposed before \cite{Scalapino1986,Monthoux1991,Monthoux1992,Moriya1990,Millis1990,Chubukov2008}, the dual fermion model not only retains their most outstanding advantage, $d$-wave pairing symmetry,  but also reconciles with the phenomenological Marginal Fermi Liquid (MFL) model \cite{Varma1989}. More importantly, it yields a Kondo-insulator-like pseudogap \cite{Aeppli1992,Coleman2007} and gets around the single-component models' difficulty that the same pairing electrons also form the pairing `glue' in superconductivity \cite{Keimer2015}. Using the hole notation, the dual fermion Hamiltonian is composed of two kinds of fermions, the doped hole quasiparticels on the O sites and localized holes on the Cu $d_{x^{2}-y^{2}}$ orbitals, 
\begin{eqnarray}
H  = H_{c}+H_{d}+J_{K}\sum_{i}\mathbf{s}_{i}\cdot \mathbf{S}_{i}+J_{H}\sum_{\left\langle i,j\right\rangle }\mathbf{S}_{i}\cdot \mathbf{S}_{j},\label{H}
\end{eqnarray}
with 
\begin{eqnarray}
H_{c}  = \varepsilon_c\sum_{i} c_{i}^{\dagger}c_{i}- \sum_{i,j}t_{ij}c_{i}^{\dagger}c_{j},  H_{d}= \varepsilon_d\sum_{i} d_{i}^{\dagger}d_{i},\label{H0}
\end{eqnarray}
where $t_{ij}$ is the hopping integral of the hole carrier quasiparticles on O sites. $c_{i}=(c_{i\uparrow},c_{i\downarrow})^T$ is the annihilation operator of carriers in unit cell $i$, and $d_{i}=(d_{i\uparrow},d_{i\downarrow})^T$ is the annihilation operator of localized holes on the Cu $d_{x^{2}-y^{2}}$ orbitals. $\mathbf{s}_{i}=c_{i}^{\dagger}\boldsymbol{\sigma}c_{i}/2$ and $\mathbf{S}_{i}=d_{i}^{\dagger}\boldsymbol{\sigma}d_{i}/2$ with the Pauli vector $\bm{\sigma}$. $J_{K}$ is the Kondo-like coupling between the carriers and localized holes, and $J_{H}$ is the Heisenberg interaction between the holes on the Cu square lattice ( for details, see Appendices~\ref{appmodel} ). 

Remarkably, different from the conventional Kondo-Heisenberg model, the much larger $J_{K}$ splits the carrier energy dispersion into two bands, possibly with overlaps, e.g. the  Zhang-Rice singlet and triplet channels in the atomic limit. Considering the hole doping concentration $\delta\ll1$, we only focus on the lower-energy band where the carrier and the Cu hole have opposite spin orientation in the same unit cell. Nevertheless, contrary to the conventional $t-J$ model \cite{Zhang1988}, we probe the low-energy properties by integrating out the Cu degrees of freedom $\mathrm{\mathbf{S}}$ rather than the O $c_i$ in the strange metal, pseudogap and superconductivity phases. 

\begin{figure}[t]
\centering
\includegraphics[width=1.0\columnwidth]{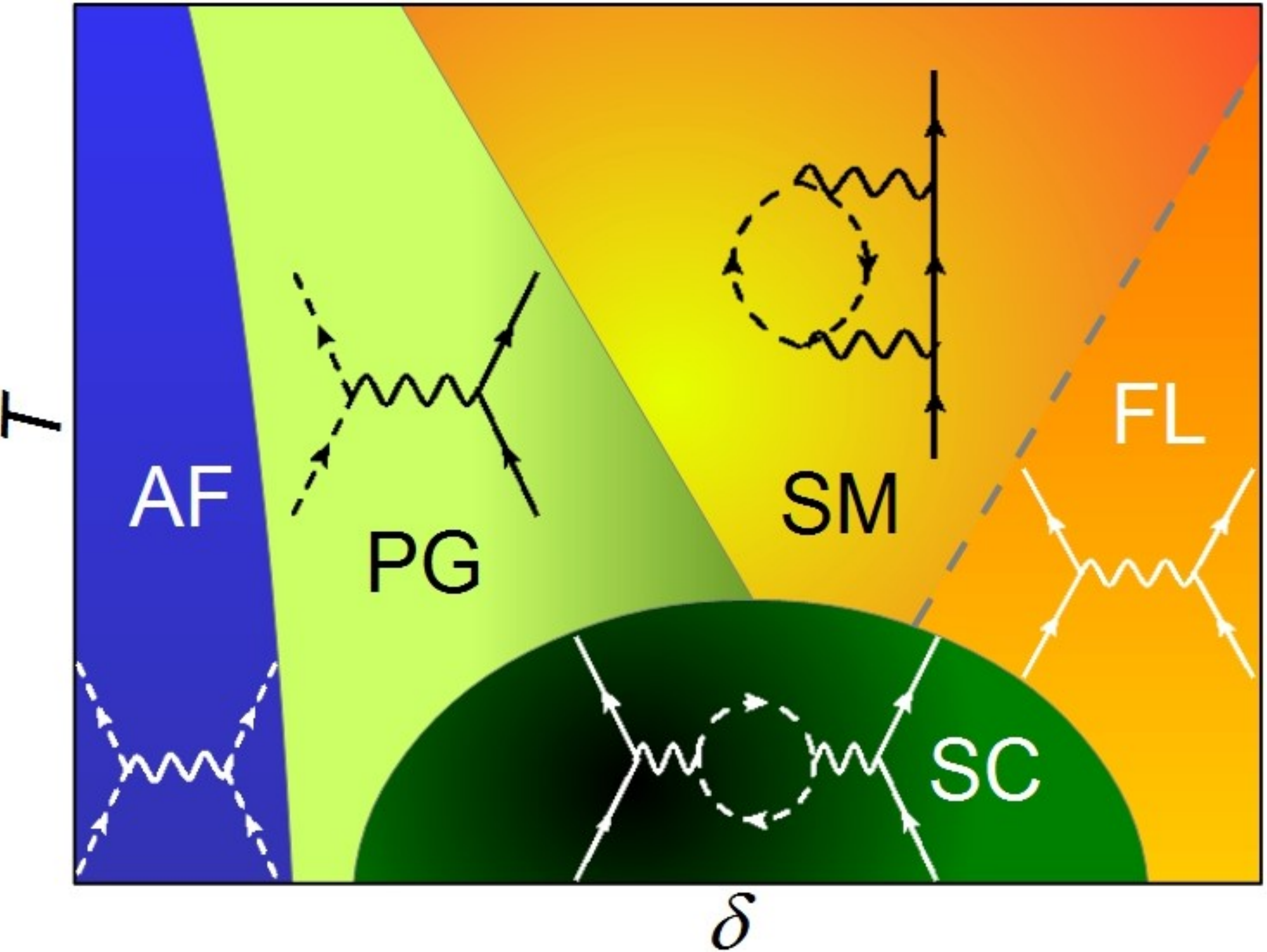}
\caption{ Schematic phase diagram of hole-doped cuprates.  Temperature $\boldsymbol{T}$ versus hole dopant concentration $\boldsymbol{\delta}$. Each regime of phase diagram is labeled with a specific interaction Feynman diagram \emph{in real space} based on the dual fermion model. The interactions are represented by the wavy lines without distinction, whereas the O and Cu holes are represented by solid and dash lines, respectively. For non-doping, the Heisenberg interaction describes the charge-transfer antiferromagnetic (AF) Mott insulators. Upon doping, antiferromagnetism gradually melts. In the strange metal (SM) phase, the doped hole Fermi gas is transformed into a non-Fermi liquid by the Heisenberg antiferromagnets. With decreasing temperature, part of the itinerant carriers couple with the local Cu magnetic moments forming localized Zhang-Rice singlets, and a pseudogap (PG) opens. In the superconductivity (SC) phase, the AF Cu-spin fluctuations are the glue of Cooper pairs. In the overdoped region, the Fermi-liquid (FL) behavior arises at low temperature.
\label{phasediagram} }
\end{figure}

In the following, we will show that the typical cuprate phase diagram can be understood solely from the Hamiltonian in Eq.~(\ref{H}). Based on this model, the characteristic interaction Feynman diagrams are plotted in each region of the phase diagram for hole-doped cuprates, as shown in Fig.~\ref{phasediagram}. In the absence of doping, only the Heisenberg interaction takes its role, describing the charge-transfer antiferromagnetic (AF) Mott insulators \cite{Anderson1987}. Upon doping, the dilute holes on the O sites are trapped around the Cu magnetic moments with opposite spin configurations, and antiferromagnetism gradually melts into a spin glass at low temperature before the onset of superconductivity with further doping. In the strange metal phase, the doped holes on O sites form a Fermi gas, coexisting with the Heisenberg antiferromagnets. Although the long range AF order disappears due to doping, the dynamical AF correlation inherits from the parent compounds. Owing to the interaction with the strongly momentum-dependent AF fluctuation, as shown in the following, the carrier Fermi gas is transformed into a non-Fermi liquid similar to the MFL. As the temperature decreases, part of the itinerant carriers bind to the local Cu magnetic moments forming localized Zhang-Rice singlets, and a pseudogap opens. In the superconductivity phase, the AF Cu-spin fluctuations mediate the conduction carriers into Cooper pairs. In the overdoped region, the system behaves as a Fermi liquid in normal state. In addition, we show that the strange metal is caused by the same underlying physics that causes superconductivity.

\section{Strange metal }
Getting inspiration from the conventional superconductivity rooted in the Fermi-liquid theory, we start from the normal state of cuprates, namely, the strange metal. Previously, although the celebrated phenomenological model of MFL can account for the universal anomalies of the strange metal \cite{Varma1989}, the proposal of $s$-wave pairing symmetry indeed dims its brilliance slightly. In particular, the experimentally confirmed $d$-wave superconductivity \cite{Wollman1993,Tsuei2000} conclusively rules out the momentum featureless fluctuations as the pairing glue. More importantly, it is baffling that why the experimentally observed strong AF spin fluctuations give ignorable contribution to the normal state, in contrast to the hypothesized momentum-uniform excitations. Additionally, the same physics is expected to cause both the strange metal and superconductivity \cite{Anderson2006}.

Here, we provide a distinct physical picture for the strange metal, in which the universal anomalies result from the strong AF fluctuations. The carriers on O sites with not heavy effective masses are assumed to compose a Fermi gas, with the retarded Green's function $G_{c}^{0}(\mathbf{k},\omega)$ since the high order interaction between the dilute carriers are ignored except in the superconductivity and Fermi-liquid phases. The Fermi gas baths in the Cu Heisenberg antiferromagnets. Most of the properties in the normal state are supposed to be determined by the retarded one-particle self-energy of the carriers due to the interaction with the Cu-spin fluctuation $\chi_{d}(\mathbf{q},\omega)$. We have omitted the first order self-energy correction, since it can be absorbed into the renormalized chemical potential. Within the Born approximation, the imaginary part of the self-energy by the spin correlation scattering reads (see Appendices~\ref{appselfenergy}): 

\begin{eqnarray}
&&\mbox{Im} \Sigma_{c}(\mathbf{k},\omega) = \frac{3J_{K}^{2}}{8\pi}\nonumber\\
&&\times \int^{\omega_c}_{-\omega_c} {dv}\left[n_{B}(v)+n_{F}(\omega+v)\right] I(\mathbf{k},\omega,v), 
\label{selfenergy}
\end{eqnarray} 
where $n_B$ and $n_F$ are the Bose and Fermi functions, $\omega_c$ is the upper cutoff frequency of magnetic fluctuations and
\begin{eqnarray}
 I(\mathbf{k},\omega,v) = \int\frac{d^2q}{4\pi^2} \mbox{Im}\chi_{d}(\mathbf{q},v)\mbox{Im}G_{c}^0(\mathbf{k+q},\omega+v),
\label{Ikwv}
 \end{eqnarray} 
where the momentum integration is over the Brillouin zone. Given that $\mbox{Im}\chi_{d}$ and $\mbox{Im}G_{c}$, corresponding to spectra functions, are continuous, integrable and do not change sign in the Brillouin zone, the second mean value theorem for integrals implies the existence of a wave vector  $\mathbf{q^*}$ as a function of $\mathbf{k}$, $v$ and $\omega$ such that
\begin{eqnarray} 
I(\mathbf{k},\omega,v) =-\pi \rho_c^0(\omega+v)\mbox{Im}\chi_{d}(\mathbf{q^*-k},v), \label{I1}
 \end{eqnarray} 
where $\rho_c^0$ is the density of states, a constant value for an ideal two-dimentional Fermi gas with quadratic dispersion. Experiments have confirmed that the magnetic susceptibility in momentum space becomes sharply peaked around AF wave vector $\mathbf{Q}$ in cuprates \cite{Tranquada1992,Hayden2004,Tranquada2004,Hinkov2007,Vignolle2007,Lipscombe2007,Fauque2007,Tacon2011,Dean2013}, which hints that $\mathbf{q^*-k} \approx \mathbf{Q}$.

For AF dynamical correlations, there is no well-developed theory or any fundamentally perturbative approach. Therefore, it is reasonable to take the imaginary part of the renormalized susceptibility from the experimentally measured spin-fluctuation spectra instead \cite{Hayden1991,Keimer1991,Tranquada1992} 
\begin{eqnarray}
\mbox{Im}\chi_{d}(\mathbf Q^*,v)\sim\mbox{Im}\chi_{d}(v)\sim \arctan\left[a_{1}\frac{v}{T}+\cdots \right],
\end{eqnarray} 
where $\mathbf Q^*$ is close to the AF wave vector $\mathbf Q$, the constant $a_1\sim1/2$ and
\begin{eqnarray}
\mbox{Im}\chi_{d}(v)\equiv \int \frac{d^2q}{4\pi^2} \mbox{Im}\chi_{d}(\mathbf{q},v).
\end{eqnarray} 
$\mbox{Im}\chi_{d}(\omega)$ becomes flat up to a cutoff energy, around 0.3 eV, according to the optical and element-specific X-ray measurements \cite{Thomas1988,Norman2006,Tacon2011,Dean2013}. Here, magnetism in cuprates is assumed to be dominated by the Cu ions. 

Substituting the experimental spin-fluctuation spectra into Eq.~(\ref{I1}) and using Eq.~(\ref{selfenergy}), we analytically calculate  the imaginary part of the Fermi gas self-energy (see Appendices~\ref{appselfenergy})
\begin{eqnarray}
\mbox{Im}\Sigma_{c}(\mathbf{k},\omega,T)&\sim& \pi \rho_c^0J_{K}^{2} T\left[1+\frac{\omega}{2T}\mbox{tanh}\left(\frac{\omega}{2T}\right)\right] \nonumber\\&\sim& \max{(|\omega|,T)}. \label{selfenergy2}
\end{eqnarray} 

More generally, integrating over the momentum $\mathbf{k}$, one finds the momentum-average imaginary part of the Fermi gas self-energy better follows the right-hand side of the Eq.~(\ref{selfenergy2}) with the aid of
\begin{eqnarray} 
\int\frac{d^{2}k}{4\pi^{2}}I(\mathbf{k},\omega,v)=-\pi\rho_{c}^{0}(\omega+v)\mbox{Im}\chi_{d}(v).
\label{I2}
 \end{eqnarray} 
For example, the momentum $\mathbf{k}$ integration of $\Sigma_{c}(\mathbf{k}, \omega=0, T)$ is actually the average over the Fermi surface with $|\mathbf{k}|=k_F$.

Clearly, a MFL-like self-energy is achieved. Therefore, the anomalies in electrical resistivity, tunneling conductance, specific heat, thermal conductivity, photoemission and the Drude contribution $\sigma_1$ in optical conductivity could be justified as in the MFL model \cite{Varma1989}. Interestingly, in the MFL model, both the nuclear relation rate and optical conductivity are assumed to result from two important contributions. Naturally, in our dual model, the two terms originate from the carriers and the Cu ions, respectively. The carriers yield the normal term proportional to $T$ in spin-lattice relaxation $^{63}T^{-1}_1$ and the Drude part in optical conductivity in spite of the non-Fermi-liquid characteristics. The Cu spin and charge fluctuations contribute the temperature-independent constant in $^{63}T^{-1}_1$ and the leading direct absorption $\sigma_2$ in the optical conductivity \cite{Thomas1988}. Corroboration is that $T^{-1}_1$ for O basically follows a Korringa rate, as in a conventional metal \cite{Varma1989,Takigawa1991}. In addition, nearly frequency- and temperature-independent, the experimentally observed featureless Raman intensity implies that the charge susceptibility $\mbox{Im}P_{d}(0,\omega)$ is proportional to $\arctan (\omega/2T)$ provided the leading inelastic Raman scattering is contributed by the Cu ions. 

\section{Pseudogap } 
Emerging from the strange metal, a partial energy gap or pseudogap opens at the Fermi level  \cite{Warren1989,Alloul1989,Norman2005,Vishik2012}, deep enough for states to become localized. Its microscopic mechanism is still puzzling. Here, based on its close association with the strange metal, we  put forward a proposal that the pseudogap derives from the dilute itinerant holes coupling to the Cu localized holes and forming localized Zhang-Rice singlets below the crossover temperature $T^*$.  

Deriving from the Kondo-like interaction in Hamiltonian Eq.~(\ref{H}) (see Appendices~\ref{appps}), the localized singlet formation process is represented by $h_{i}^{\dagger}c_i$, where the fermionic composite operator $h_i$ is introduced to denote the annihilation of a singlet accompanying the creation of a Cu $d$ hole in unit cell $i$, 
\begin{eqnarray}
h_{i\sigma} \equiv d_{i\sigma}^\dagger\frac{1}{ \sqrt2}(d_{i\uparrow} c_{i\downarrow}-d_{i\downarrow} c_{i\uparrow}). \label{}
\end{eqnarray}
The creation of a composite fermion breaks the magnetic bonds with its four surrounding Cu spins and is effectively equivalent to a soliton on the spin-$1/2$ square lattice. In addition, the assumption of a singly occupied Cu hole results in the conclusion that no more than one composite fermion resides in any unit cells. In this phase, the carriers and the composite fermions coexist and transform each other with the same chemical potential, and the Kondo-like interaction in Eq.~(\ref{H}) is rewritten as 
\begin{eqnarray}
H_{K}=-\frac{3\sqrt2 J_K}{16} \sum_{i\sigma}(\sigma c_{i\sigma}^{\dagger}h_{i\bar\sigma}+H.c.)\label{HK}
\end{eqnarray}
in the singlet channel, where $\sigma=\pm1$ corresponding to spin up and down, respectively. It is straightforward that a gap opens in the system due to the hybridization between the carriers and the dispersionless composite quasiparticles in Eq.~(\ref{HK}), just as in Kondo insulators \cite{Aeppli1992,Coleman2007}. The indirect `hybridization gap' $\Delta_{ind}$ is around $J_K^2/7D$, much smaller than the direct gap $\Delta_{dir}=3\sqrt2 J_K/8$, where $D$ is the half-width of the conduction band \cite{Coleman2007}, and $\Delta_{ind}\sim J_H$ (ref. \cite{Zaanen1988}). Considering the doping dependence of the effective Heisenberg coupling \cite{Barzykin2009}, the indirect gap should be proportional to $J_H(1-a\delta)$ with  a constant $a$ and  the hole doping concentration $\delta$. Moreover, in this hybridization picture, the total particle number is conserved during the transformation between the carriers and the localized composite quasiparticles, i.e. $x_c+x_h=\delta$, where $x_c$ and $x_h$ are the concentrations of the carriers and the composite quasiparticles, respectively. With decreasing of temperature or doping level, a higher proportion of the doped holes bind to Cu local moments into the localized singlets. Particularly, in AF insulator phase, no carriers survive, namely, $x_c=0$. 

The presence of an energy gap often induces the increase of the resistance but the formation of localized Zhang-Rice singlets reduces the AF fluctuation scattering rate on the Cu-O planes. This competition potentially leads to the drop of planar resistivity in contrast with the upturn of the c-axis resistance for temperatures below $T^*$ (refs. \cite{Puchkov1996,Homes1993}).  Actually, the cuprates with anisotropic pseudogaps are more akin to nodal Kondo insulators. Nevertheless, the nodal character and doping dependence of Kondo insulators are poorly understood from the simple hybridization picture \cite{Coleman2007}. It is also worth noting that different from the conventional Kondo lattice, the singlets in pseudogap are randomly scattered across the lattice rather than over each site especially in the underdoped regime. Moreover, the superconductivity fluctuation, phonons, Fermi surface nesting and the Coulomb interaction between charges further complicate the situations, e.g. the formation of preformed Cooper pairs, stripe phase, nematic order, phase separation, spin and/or charge density order \cite{Norman2005,Vishik2012,Doiron-Leyraud2013,Keimer2015,Badoux2016}.  The charge density wave order presented in many cuprate families could be described analogously based on the hot-spot mechanism as in the spin-fermion model \cite{Wang2014}.

\section{Superconductivity}
To get insight into the origin of superconductivity emerging from the pseudogap and/or  strange metal, we propose that the carrier pairing is mediated by the high-energy short-range Cu-spin fluctuations, persisting into the superconducting phase from the normal state. This mechanism is analogous to the magnetic-exciton pairing mechanism \cite{McHale2004} in heavy fermion compound UPd$_2$Al$_3$. As described above, the isolated Zhang-Rice singlets break their surrounding AF bonds and then the localized Cu holes composing the singlets are completely separated from other local moments. On the other hand, the local moments may only be partially screened by doped holes without formation of onsite singlets, and then the Heisenberg interaction between the screened moments and their surroundings could survive. Thus, the carriers on unit cell $i$ and $j$ could interact with each other by exchanging the Cu-spin fluctuations. Since the energy scale of magnetic interactions is comparable with the carrier bandwidth, so all momenta are involved in magnetic interactions \cite{Coleman2007}. The interaction Hamiltonian can be written in the coordinate representation as (also shown in Fig.~\ref{phasediagram} and  Appendices~\ref{appsc})
\begin{eqnarray}
H_{sc}  =J_{K}^{2} \sum_{ij}  \chi_{d}(i,j,\omega)\mathbf{s}_{i}\cdot \mathbf{s}_{j}.\label{Hsc}
\end{eqnarray}
Combining with $H_{c}$ in Eq.~(\ref{H0}), a $t$-$J$-like model is reached. Interestingly, despite the formal similarity with the conventional $t$-$J$ model \cite{Zhang1988}, here the spin-like operator $\mathbf{s}$ is associated with the carriers on the O sites rather than the local moments on Cu ions. Therefore, there is no singly-occupied constraint on the carrier sites, and thus it is distinct from the conventional  $t$-$J$ model.

Since the magnetic susceptibility in momentum space sharply peaks around AF wave vector $\mathbf{Q}$ in cuprates \cite{Tranquada1992,Hayden2004,Tranquada2004,Hinkov2007,Vignolle2007,Lipscombe2007,Fauque2007,Tacon2011,Dean2013}, only the $d$-wave pairing channel is favored \cite{Scalapino1986,Bickers1987,Inui1988,Dong1988,Kotliar1988,Monthoux1991,Monthoux1992,Moriya1990,Millis1990,Chubukov2008} and the attractive pairing interaction dominantly mediates the carriers on the nearest-neighbor unit cells \cite{Scalapino2012}. In consideration of the two lattice spacings order of the AF spin correlation length at low temperature, we only take the nearest-neighbor coupling into account in Eq.~(\ref{Hsc}) and define an effective coupling between the carriers, \begin{eqnarray}
g\equiv \frac{3}{4}J_K^2\chi_{d}(\left\langle i,j \right\rangle,\omega).
\end{eqnarray}
$g$ is approximately assumed to be a constant within the spin fluctuation cutoff energy $\hbar \omega_c$, otherwise, $0$. Consequently, in momentum space, a weak coupling BCS interaction can be written in $d$-wave pairing channel
\begin{eqnarray}
H_{scd}=-g\int  \frac{d^2 k}{4\pi^2} \gamma_{\mathbf{k}}c_{\mathbf{k}\uparrow}^{\dagger}c_{-\mathbf{k}\downarrow}^{\dagger}  \int \frac{d^2 k'}{4\pi^2}\gamma_{\mathbf{k'}}c_{\mathbf{k'}\uparrow}c_{-\mathbf{k'}\downarrow} ,
\end{eqnarray}
where $\gamma_\mathbf{k}=(\cos k_x-\cos k_y)$ is the $d_{x^2-y^2}$-wave gap function.
The superconductivity order parameter $\gamma_{\mathbf{k}}\Delta_{sc}$ is introduced in the mean field method with the BCS gap equation
\begin{eqnarray}
\Delta_{sc}=-g \int\frac{d^2 k}{4\pi^2}\gamma_{\mathbf{k}}<c_{\mathbf{k}\uparrow}^{\dagger}c_{-\mathbf{k}\downarrow}^{\dagger}>,
\end{eqnarray}
where $c_{\mathbf{k}\uparrow}^{\dagger}c_{-\mathbf{k}\downarrow}^{\dagger}$ is the Cooper pair operator denoting a bond state of two carriers with opposite momentum and spin. 

Intuitively, the superconductivity gap should depend on the density of superconducting carriers. However, this dependence is concealed in the formula of the zero temperature energy gap $\Delta_{sc}$ and the critical temperature $T_c$ in the BCS theory. The concealment results from the implicit BCS assumption that the pairing cutoff energy or  Debye energy is far less than the electron Fermi energy or chemical potential $E_F$, namely, the Migdal's theorem, being perfect for metals. However, in lightly doped cuprates, the doped hole Fermi energy $E_F\approx2D\delta$ could be less than the relatively high spin fluctuation cutoff energy $\hbar\omega_c$. Despite the violation of the Migdal's limits, it is generally believed that the cuprate superconductivity is basically BCS-like \cite{Matsui2003,Scalapino2012,Keimer2015}. In this case, we propose that the integral cutoff energy in the BCS gap equation should be $\min({E_F ,\hbar\omega_c})$ since there exists no carrier state below the bottom of the energy band (see Appendices~\ref{appgapequ}). To some extent, this revision restores the dependence of the gap on the superconducting carrier density. For instance, when the Fermi energy $E_F$ is less than the spin fluctuation cutoff $\hbar\omega_c$, the superconductivity transition temperature $T_c\sim \delta D e^{-1/\lambda}$, otherwise $T_c\sim \omega_c e^{-1/\lambda}$ with $\lambda=g\rho^0_{c}/2$ for weak coupling $d$-wave superconductors. It is worthy of noting that the coupling $g$ associated with spin fluctuations should be gradually suppressed with doping increasing \cite{Chia2013}. Consequently, the characteristic dome-shaped $T_c$ versus doping level is sketched out for hole-doped cuprates. To further underline the importance of the revised cutoff energy in the gap equation, assuming the Fermi energy $2\delta D$ less than $\hbar\omega_c$, then we find $\delta<0.15$, ranging almost from the zero doping to optimal doping, where we have set the half-bandwidth $D=$1 eV and $\hbar\omega_c=0.3$ eV. It is still premature to quantitatively calculate the critical temperature $T_c\sim \omega_c e^{-1/\lambda}$ because it is sensitive to the coupling $\lambda$ due to the large spin fluctuation cutoff energy, for example, $\lambda=0.2-0.3$, then $T_c\approx27$ K$-141$ K. In addition, the coupling $g$ is difficult to calculate quantitatively and the Coulomb repulsion between pairing carriers has been ignored. 

Finally, it is noteworthy that in our model, the Cu-spin-fluctuation frequency for pairing is much higher than the characteristic frequency in one-component models \cite{Monthoux1991,Moriya1990,Millis1990,Chubukov2008}. Nevertheless, the higher pairing frequency is reconcilable with the experimental measurements \cite{Hayden2004,Tranquada2004,Hinkov2007,Vignolle2007,Lipscombe2007,Fauque2007,Tacon2011,Dean2013} and the existing theoretical conclusions that low-frequency spin fluctuations are pair breaking \cite{Millis1988,Anderson2007} and the optimal spectral weight for pairing originates from a frequency range larger than twice the maximum value of the superconducting gap \cite{Monthoux1994}.

\section{Discussion}
Despite using a hole representation in the model Hamiltonian, our results are independent of the hole or electron notation. The concentration of the spin unpaired holes and electrons on O sites are always the same, corresponding to the hole doping level $\delta$. Only a singly occupied hole or electron on an O site couples to a localized Cu hole composing a Zhang-Rice singlet in our model. Similarly, only the spin unpaired holes or electrons on the O sites are mediated by the Cu-spin fluctuations via the Kondo-like interactions. 

The Fermi surface is crucial in hole-doped cuprates although their normal states are non-Fermi liquids. As in the Kondo lattice model, Luttingers theorem survives \cite{Martin1982,Oshikawa2000}. However, the screened localized holes also contribute to the Fermi surface volume. In the strange metal phase, we tend to apply the large Fermi surface with an area corresponding to $1+\delta$ holes in despite of the marginal Fermi liquid \cite{Norman1998}. In the underdoped regime, besides the itinerant and localized holes, the massless composite fermions arose below $T^*$, analogous to the Cooper pairs formation under $T_c$. It needs to be understood whether the composite fermion involvement leads to the Fermi surface shrinking down into the Fermi arcs. 

Markedly, it is desired to understand the AF dynamical correlations and justify the neglect of vertex corrections in both self-energy and the pairing interaction in this work beyond the arguments \cite{Chubukov2008}. The ignored intra-unit-cell freedoms should endow more detailed characteristics \cite{Simon2002}. Much more substantial work is needed to clarify the extremely complex phenomena, such as the nodal character and doping dependence in pseudogap, space and/or time symmetry breaking, the strong coupling superconductivity effects, the Fermi arcs and the Lifshitz transition of Fermi surface \cite{Lee2006,Doiron-Leyraud2013,Badoux2016,Keimer2015}. To test the dual fermion model, it is of crucial importance to separate the Cu-spin fluctuations from the carrier magnetic excitations. We propose that the element-specific experimental measurements are competent, such as inelastic X-ray scattering and NMR. Actually, our dual Fermi model is in good agreement with the existing NMR observations where the magnetic susceptibility must be decomposed into two component contributions from Cu and O, respectively \cite{Haase2009}. In the pseudogap phase, we expect that the direct Kondo-insulator-like gap could be visible to optical conductivity and the ratio of carrier to localized composite fermion could be detected in the carrier-sensitive Hall experiments. The validity of our generalized BCS gap equation could be examined in the superconductors with low superfluid density. 

In conclusion, the itinerant-localized dual Fermi model can unify the main universal features of the hole-doped cuprates. Our work paves a novel way to the studies of both non-Fermi liquid and unconventional superconductivity in strongly correlated systems.

\section{Acknowledgements}
We are thankful to S. J. Qin, T. Xiang and J. Y. Gan for fruitful discussions.

\appendix
\begin{center}
{\textbf{Supplemental Material for: \\ \textit{Theory of dual fermion superconductivity in hole-doped cuprates}}}
\end{center}

In this Supplemental Material we present the details on the deduction of the dual fermion model, self-energy in strange metal phase, the pseudogap formation and the d-wave superconductivity mechanism.
\setcounter{equation}{0}
\renewcommand{\theequation}{A\arabic{equation}}

\section{\label{appmodel}Dual fermion model}

The parent compound of cuprate superconductivity is a Mott insulator.
Each Cu ion has one hole and all the O sites are empty. Virtual hopping
processes lead to antiferromagnetic superexchange between the Cu spins.
Upon doping, additional holes are introduced on the Cu-O planes, dominantly
on O sites. Emery proposed a three-band model to describe a single
layer of Cu-O plane in the hole representation \cite{Emery1987},

\begin{eqnarray}
H_{3B}=&&H_{pd}+\sum_{i\sigma}\varepsilon_{d}^{0}n_{di\sigma}+\sum_{l\sigma}\varepsilon_{p}^{0}n_{pl\sigma}\nonumber\\
&+&U_{dd}\sum_{i}n_{di\uparrow}n_{di\downarrow}+U_{pp}\sum_{l}n_{pl\uparrow}n_{pl\downarrow},
\end{eqnarray}
where $n_{di}$ and $n_{pl}$ are the particle number operators of
Cu holes at site $i$ and O holes at site $l$, respecitvely. $\varepsilon_{d}^{0}$
and $\varepsilon_{p}^{0}$ denote the energy levels occupied by holes
on the Cu sites and O sites. $U_{dd}$ and $U_{pp}$ are the onsite
Coulomb interaction of Cu and O holes, respectively. The hybridization
between Cu and O holes is given by 
\begin{eqnarray}
H_{pd}=-\sum_{<i,l>\sigma}(t_{il}d_{i\sigma}^{\dagger}p_{l\sigma}+H.c.),
\end{eqnarray}
where $t_{il}=(-1)^{M_{il}}t_{pd}$ are the overlaps of the corresponding
orbitals with the amplitude of the hybridization $t_{pd}$. $M_{il}=2$
if $\mathbf{R}_{l}=\mathbf{R}_{i}-\mathbf{x}/2$ or $\mathbf{R}_{i}-\mathbf{y}/2$
and $M_{il}=1$ if $\mathbf{R}_{l}=\mathbf{R}_{i}+\mathbf{x}/2$ or
$\mathbf{R}_{i}+\mathbf{y}/2$, where $\mathbf{R}_{i}$ represents
the position vector of the $i$th Cu atom and $\mathbf{R}_{l}$ is
the position vector of its nearest-neighbouring O atoms, and $\mathbf{x}$
and $\mathbf{y}$ denote the unit vectors along the x and y directions,
respectively.

It is customary to further simplify this model. We shall consider
the case $U_{pp}=0$, $t_{pd}\ll U_{dd}$ and $t_{pd}\ll\Delta$,
with the charge transfer energy $\Delta=\varepsilon_{p}-\varepsilon_{d}$.
A common way is to use the limit $U_{dd}\rightarrow\infty$ to forbid
double occupancy in the $3d$ orbitals, i.e. $n_{di}=1$. We define
a projection operator $\hat{P}_{g}$, which projects on the subspace
with only one hole locating on each Cu site, and another projector
$\hat{P}_{e}=\hat{I}-\hat{P}_{g}$ with the unit operator $\hat{I}$.

We regard the $H_{pd}$ term as the perturbation to derive a effective
Hamiltonian. For more technical details we refer to refs.
\cite{Zaanen1988,Zhang1988,Ogata1988,Shen1990}. One finds the first and the
third order perturbations vanish,

\begin{eqnarray}
H^{(1)}=\hat{P}_{g}H_{pd}\hat{P}_{g}=0,
\end{eqnarray}
and 
\begin{eqnarray}
H^{(3)}=\hat{P}_{g}H_{pd}(\frac{1}{E-H_{3B}}\hat{P}_{e}H_{pd})^{2}\hat{P}_{g}=0.
\end{eqnarray}

The second-order contribution reads

\begin{eqnarray}
H^{(2)}=\hat{P}_{g}H_{pd}(\frac{1}{E-H_{3B}}\hat{P}_{e}H_{pd})\hat{P}_{g}.
\end{eqnarray}
The detail result is written as 
\begin{eqnarray}
H^{(2)}&&=-\sum_{<il><im>\sigma\sigma^{\prime}}\hat{P}_{g}t_{im}t_{il}\times \nonumber \\
&&\left(\frac{d_{i\sigma}^{\dagger}d_{i\sigma^{\prime}}p_{l\sigma}p_{m\sigma^{\prime}}^{\dagger}}{\Delta}+\frac{d_{i\sigma}d_{i\sigma^{\prime}}^{\dagger}p_{l\sigma}^{\dagger}p_{m\sigma^{\prime}}}{U_{dd}-\Delta}\right)\hat{P}_{g}.
\end{eqnarray}

To combine the four oxygen-hole states around a Cu ion, operators
$P_{i\sigma}$ and $P_{i\sigma}^{A}$ are introduced \cite{Zhang1988} 
\begin{eqnarray}
P_{i\sigma}\equiv\frac{1}{2}\sum_{<i,l>}(-1)^{M_{il}}p_{l\sigma}
\end{eqnarray}
and 
\begin{eqnarray}
P_{i\sigma}^{A}\equiv\frac{1}{2}\sum_{<i,l>}p_{l\sigma},
\end{eqnarray}
where the sum runs over the four O sites $l$ around a given Cu site
$i$. Accordingly, we introduce the momentum-space operators $P_{\mathbf{k}\sigma}$
and $P_{\mathbf{k}\sigma}^{A}$ 
\begin{eqnarray}
P_{i\sigma}\equiv\int\frac{d^{2}k}{4\pi^{2}}\beta_{k}^{-1}P_{\mathbf{k}\sigma}e^{i\mathbf{k}\cdot\mathrm{\mathbf{R}_{i}}}
\end{eqnarray}
and 
\begin{eqnarray}
P_{i\sigma}^{A}\equiv\int\frac{d^{2}k}{4\pi^{2}}\beta_{k}^{-1}P_{\mathbf{k}\sigma}^{A}e^{i\mathbf{k}\cdot\mathbf{\mathrm{\mathbf{R}_{i}}}},
\end{eqnarray}
with $\beta_{k}=\sqrt{1-(\cos k_{x}+\cos k_{y})/2}$, and the momentum
integration is over the Brillouin zone.

Then the $H^{(2)}$ is expressed as 
\begin{eqnarray}
H^{(2)}&&=-4\frac{t_{pd}^{2}}{\Delta}\sum_{i}\hat{P}_{g}n_{di}\hat{P}_{g}+4\frac{t_{pd}^{2}}{\Delta}\sum_{i}\hat{P}_{g}P_{i\sigma}^{\dagger}P_{i\sigma}\hat{P}_{g}+J_{dP}\times\nonumber \\
&&\sum_{i}\hat{P}_{g}\left(d_{i\sigma}^{\dagger}d_{i-\sigma}P_{i-\sigma}^{\dagger}P_{i\sigma}-d_{i\sigma}^{\dagger}d_{i\sigma}P_{i-\sigma}^{\dagger}P_{i-\sigma}\right)\hat{P}_{g},
\end{eqnarray}
with $J_{dP}=4t_{pd}^{2}\left(\frac{1}{\Delta}+\frac{1}{U_{dd}-\Delta}\right)$.

Due to two neigbouring unit cells sharing the same O atom, $P_{i\sigma}$
are normalized but not orthogonalized. Therefore, Wannier operators
at $\mathbf{\mathrm{\mathbf{R}}}_{i}$ are introduced \cite{Zhang1988}

\begin{eqnarray}
c_{i\sigma}\equiv\int\frac{d^{2}k}{4\pi^{2}}P_{\mathbf{k}\sigma}e^{i\mathbf{k}\cdot\mathbf{\mathrm{\mathbf{R}}}_{i}}
\end{eqnarray}
and 
\begin{eqnarray}
b_{i\sigma}\equiv\int\frac{d^{2}k}{4\pi^{2}}P_{\mathbf{k}\sigma}^{A}e^{i\mathbf{k}\cdot\mathbf{\mathrm{\mathbf{R}}}_{i}},
\end{eqnarray}
where the operators $c_{i}$ and $b_{i}$ are orthogonalized and complete
in the O-hole space, and
\begin{eqnarray}
\sum_{l\sigma}\varepsilon_{p}^{0}n_{pl\sigma}=\sum_{i\sigma}\varepsilon_{p}^{0}\left(c_{i\sigma}^{\dagger}c_{i\sigma}+b_{i\sigma}^{\dagger}b_{i\sigma}\right).
\end{eqnarray}
Only $c_{i}$ couples to the local moments, so that the $b_{i}$ states
are nonbonding states. 

Considering terms up to the nearest-neighboring hopping process in
the Wannier representation, and only keeping the Cu-O interaction
terms in the same unit cells, then $H^{(2)}$ can be rewritten as
\begin{eqnarray}
H^{(2)}=&-&4t_{1}\sum_{i}\hat{P}_{g}n_{di}\hat{P}_{g}\nonumber \\
&+&J_{K}\sum_{i}\hat{P}_{g}\left(\mathrm{\mathbf{S}}_{i}\cdot\mathrm{\mathbf{s}}_{i}-\frac{1}{4}n_{di}n_{ci}\right)\hat{P}_{g}\nonumber \\&+&4t_{1}u^{2}(0,0)\sum_{i}\hat{P}_{g}n_{ci}\hat{P}_{g}\nonumber \\
&+&8t_{1}u(0,0)u(0,1)\sum_{\left\langle ij\right\rangle }\hat{P}_{g}c_{j\sigma}^{\dagger}c{}_{i\sigma}\hat{P}_{g},
\end{eqnarray}
where $t_{1}=t_{pd}^{2}/\Delta$, and the coupling between the carriers
and local moments is 
\begin{eqnarray}
J_{K}=8\left(t_{1}+t_{2}\right)u^{2}(0,0)
\end{eqnarray}
with $t_{2}=t_{pd}^{2}/\left(U_{dd}-\Delta\right)$, and 
\begin{eqnarray}
u(i,j)=\int\frac{d^{2}k}{4\pi^{2}}\beta_{k}e^{i\mathbf{k}\cdot\left(\mathbf{\mathrm{\mathbf{R}}}_{i}-\mathbf{\mathrm{\mathbf{R}}}_{j}\right)}.
\end{eqnarray}
One finds $u(i,i)\approx0.96$ and $u(i,i+1)\approx-0.14$.

The fourth-order perturbation is written as

\begin{eqnarray}
H^{(4)}=\hat{P}_{g}H_{pd}(\frac{1}{E-H_{3B}}\hat{P}_{e}H_{pd})^{3}\hat{P}_{g}.
\end{eqnarray}
Out of many fourth order terms, the Cu-Cu Heisenberg superexchange
plays an important role on the square lattice 
\begin{eqnarray}
H^{(4)}\approx J_{H}\sum_{<i,j>}\hat{P}_{g}\mathrm{\mathbf{S}}_{i}\cdot\mathrm{\mathbf{S}}_{j}\hat{P}_{g},
\end{eqnarray}
with $J_{H}=4t_{pd}^{4}\left(\frac{1}{\Delta^{3}}+\frac{1}{\Delta^{2}U_{dd}}\right)$,
and we have ignored the hopping terms.

Finally, casting off constants and the nonbonding states $b_{i}^{\dagger}b_{i}$,
an effective Hamiltonian up to the fourth-order perturbation is reached

\begin{eqnarray}
H_{eff}=&&\sum_{i}\varepsilon{}_{d}n_{di}+\sum_{i}\epsilon{}_{p}n_{ci}-\sum_{ij\sigma}t_{ij}c_{i\sigma}^{\dagger}c_{j\sigma}\nonumber \\&+&J_{K}\sum_{i}\mathrm{\mathbf{s}}_{i}\cdot\mathbf{S}_{i}+J_{H}\sum_{<i,j>}\mathrm{\mathbf{S}}_{i}\cdot\mathrm{\mathbf{S}}_{j},
\end{eqnarray}
where $t_{ij}$ describe hoping terms, $\epsilon{}_{d}=\epsilon_{d}^{0}-4t_{1}$
and $\epsilon{}_{p}=\epsilon_{p}^{0}+4t_{1}u^{2}(0)-J_{K}/4$. The
projection operators have been omitted. Moreover, the O-O hopping
between unit cells could be included in $t_{ij}$.

In the atomic limit, the AF coupling between an O hole and a localized Cu hole forms the singlet and triplet energy levels. The gap between the two energy levels is proportional to $J_{K}$. In the lattice system, the two levels are broadened into two bands, potentially
with overlaps. If we further project the Hamiltonian to the Zhang-Rice singlet
subspace, then we obtain the celebrated $t$-$J$ model \cite{Zhang1988}.
However, in our work, contrary to the $t$-$J$ model, we integrate
out the Cu degrees of freedom $\mathrm{\mathbf{S}}$ rather than the
O $c_{i}$ in the lower energy band. Importantly, as a carrier fails
to completely screen a local moment, we take into account the Cu-spin
fluctuations effects on the carrier, which give substantial contribution
to the carrier self-energy and Cooper pairing interaction. 

\setcounter{equation}{0}
\renewcommand{\theequation}{B\arabic{equation}}
\section{\label{appselfenergy}MFL-like self-energy}

The generic characteristics in Fermi-liquid metals are attributed
to the scattering between fermions. However, in the metallic state
of high temperature cuprate superconductors this theory fails in the
presence of strong magnetic correlations. We ascribe the anomalies
in normal state to the dominant interaction between the fermions and
the Cu-spin fluctuations rather than the coupling between fermions.
We assume that the unrenormalized Wannier quasiparticles or hole carriers
form a Fermi gas on the Cu-O plane. The anomalous transport and thermodynamic
properties dominantly hinge on the carrier self-energy renormalized
by the scattering of Cu-spin fluctuations 
\begin{eqnarray}
&&\Sigma{}_{c\alpha}(\mathbf{k},\omega_{n})=-iT\sum_{v_{n}\beta}J_{K}^{2}\sigma_{\alpha\beta}\cdot\sigma_{\beta\alpha} \times \nonumber\\ &&\int\frac{d^{2}q}{4\pi^{2}}G_{c\beta}^{0}(\mathbf{k}+\mathbf{q},\omega_{n}+v_{n})\chi_{d}(\mathbf{q},v_{n}).
\end{eqnarray}
with Matsabra frequency $v_{n}=2\pi nT$. The production of the Pauli
matrix can be decomposed into triplet and singlet spin configurations,
respectively\cite{Chubukov2008,Coleman2007}, 
\begin{eqnarray}
\sigma_{\alpha\beta}\cdot\sigma_{\gamma\delta}=\frac{1}{2}\left(\delta_{\alpha\beta}\delta_{\gamma\delta}+\delta_{\alpha\delta}\delta_{\beta\gamma}\right)\nonumber \\
-\frac{3}{2}\left(\delta_{\alpha\beta}\delta_{\gamma\delta}-\delta_{\alpha\delta}\delta_{\beta\gamma}\right). \label{PTS}
\end{eqnarray}
From the real space point of view, for a carrier hopping on a spin-$1/2$ square lattice, its spin orientation is opposite to that of the Cu hole in the same unit cell with the carrier. Performing analytic continuation and ignoring the spin index, we obtain
\begin{eqnarray}
&&\mbox{Im}\Sigma{}_{c}(\mathbf{k},\omega)\sim J_{K}^{2}\nonumber \\ &&\times\int_{-\omega_{c}}^{\omega_{c}}\frac{dv}{2\pi} \left[n_{B}(v)+n_{F}(\omega+v)\right]I(\mathbf{k},\omega,v) \label{ase},
\end{eqnarray}
with the spin-fluctuation cutoff frequency $\omega_{c}$ and 
\begin{eqnarray}
I(\mathbf{k},\omega,v)=\int\frac{d^{2}q}{4\pi^{2}}\mbox{Im}G_{c}^{0}(\mathbf{k+q},\omega+v)\mbox{Im}\chi_{d}(\mathbf{q},v).
\end{eqnarray}

Given that $\mbox{Im}\chi_{d}$ and $\mbox{Im}G_{c}$, corresponding to spectra functions, are continuous, integrable and do not change sign in the Brillouin zone, the second mean value theorem for integrals implies the existence of a wave vector  $\mathbf{q^*}$ as a function of $\mathbf{k}$, $v$ and $\omega$ such that
\begin{eqnarray} 
I(\mathbf{k},\omega,v) =-\pi \rho_c^0(\omega+v)\mbox{Im}\chi_{d}(\mathbf{q^*-k},v), \label{aI}
 \end{eqnarray} 
where $\rho_c^0$ is the density of states, a constant value for an ideal two-dimentional Fermi gas with quadratic dispersion. Since the magnetic susceptibility in momentum space becomes sharply peaked around AF wave vector $\mathbf{Q}$ in cuprates, which indicates $\mathbf{q^*-k} \approx \mathbf{Q}$.

We take the imaginary part of the renormalized susceptibility from the experimentally measured spin-fluctuation spectra instead \cite{Hayden1991,Keimer1991,Tranquada1992} 
\begin{eqnarray}
\mbox{Im}\chi_{d}(\mathbf Q^*,v)\sim e^{-\frac{(\mathbf {Q^*-Q})^2}{2\sigma^2}}\arctan\left[a_{1}\frac{v}{T}\right],
\end{eqnarray} 
where $\mathbf Q^*$ is close to the AF wave vector $\mathbf Q$, the constant $a_1\sim1/2$ and $\sigma$ is the wave vector width.

Substituting the experimental spin-fluctuation spectra into Eq.~(\ref{aI}) and using Eq.~(\ref{ase}), we get  the imaginary part of the Fermi gas self-energy
\begin{eqnarray}
&&\mbox{Im}\Sigma{}_{c}(\mathbf{k},\omega)\sim e^{-\frac{(\mathbf {\mathbf{q^*-k}-Q})^2}{2\sigma^2}}\times\nonumber\\ &&\int_{-\omega_{c}}^{\omega_{c}}\frac{dv}{2\pi}\left[n_{B}(v)+n_{F}(\omega+v)\right]\arctan\left[\frac{v}{2T}\right].
\end{eqnarray} 
where we have ignored the frequency dependence of $\sigma$ and $\mathbf{q^*}$. Since $\mathbf{q^*-k}$ is close to the AF wave vector $\mathbf{Q}$, $\mathbf{q^*}$ must be insensitive to $v$. 

Since the absolute value of $n_{B}(v)+n_{F}(\omega+v)$ exponentially
decreases to zero with increasing $|v|$, the range of integration can be extended from
the cutoff $\omega_{c}$ to infinity. Using the similar hyperbolic
tangent function $\tanh(x)$ to replace the arc tangent function $\arctan(x)$,
then the frequency integral can be calculated analytically similar
to the calculation in ref. \cite{Chubukov2012}

\begin{eqnarray}
\int_{-\infty}^{\infty}dv\left[n_{B}(v)+n_{F}(\omega+v)\right]\mbox{tanh}\left(\frac{v}{2T}\right) \nonumber\\=2T\left[1+\frac{\omega}{2T}\mbox{tanh}\left(\frac{\omega}{2T}\right)\right].
\end{eqnarray}

A marginal Fermi liquid (MFL)-like self-energy is reached 
\begin{eqnarray}
\mbox{Im}\Sigma{}_{c}(\mathbf{k},\omega)&\sim& e^{-\frac{(\mathbf {q^*-k-Q})^2}{2\sigma^2}}T\left[1+\frac{\omega}{2T}\mbox{tanh}\left(\frac{\omega}{2T}\right)\right] \nonumber\\&\sim& e^{-\frac{(\mathbf {q^*-k-Q})^2}{2\sigma^2}}\max\left(T,\left|\omega\right|\right).
\end{eqnarray}

On the other hand, without applying the approximation of weak frequency dependence of $\sigma$ and $\mathbf{q^*}$, we obtain the momentum-average imaginary part of the Fermi gas self-energy,
\begin{eqnarray}
&&\int\frac{d^{2}k}{4\pi^{2}}\mbox{Im}\Sigma{}_{c}(\mathbf{k},\omega)\sim\nonumber\\ &&\int_{-\omega_{c}}^{\omega_{c}}\frac{dv}{2\pi}\left[n_{B}(v)+n_{F}(\omega+v)\right] \int\frac{d^{2}k}{4\pi^{2}}I(\mathbf{k},\omega,v).
\end{eqnarray}
with
\begin{eqnarray}
\int\frac{d^{2}k}{4\pi^{2}}I(\mathbf{k},\omega,v)=-\pi\rho_{c}^{0}(\omega+v)\mbox{Im}\chi_{d}(v),
\end{eqnarray}
where the definition $\mbox{Im}\chi_{d}(v)\equiv\int d^{2}q\mbox{Im}\chi_{d}(\mathbf{q},v)/4\pi^{2}$
and $\rho_{c}^{0}(\omega+v)$ is the density of states of the Fermi gas, a constant value.

Substituting the imaginary part of the renormalized Cu-spin susceptibility with the experimental spin-fluctuation spectra, i.e. $\mbox{Im}\chi_{d}(v)\sim\mbox{tanh}\left(\frac{v}{2T}\right)$, we rewrite the momentum-average imaginary part of the self-enegy as
\begin{eqnarray}
\int\frac{d^{2}k}{4\pi^{2}}\mbox{Im}\Sigma{}_{c}(\mathbf{k},\omega)&&\sim \pi\rho_{c}^{0}J_{K}^{2}T\left[1+\frac{\omega}{2T}\mbox{tanh}\left(\frac{\omega}{2T}\right)\right] \nonumber\\ &&\sim \max\left(T,\left|\omega\right|\right).
\end{eqnarray}

Thus, the universal anomalies in the strange metal originate simply from the Fermi gas renormalized by strong Cu-spin fluctuations. Actually, early in 2006, the power-law optical conductivity was attributed to
a consequence of the carriers interacting with a broad spectrum of bosons \cite{Norman2006}.

\setcounter{equation}{0}
\renewcommand{\theequation}{C\arabic{equation}}
\section{\label{appps}Hybridization Hamiltonian in pseudogap}

The origin of the pseudogap is unclear so far. For example, it is
still controversial whether the gap results from the formation of
spin singlets, nematic order, spin, charge or $d$-wave of density
wave and so on. Moreover, a $d$-wave versus $s$-wave symmetry of
the gap is still debating \cite{Sakai2013,LiYuan2013}. The conclusive evidences
are the experimentally observed partial gaps in various spectroscopys,
such as nuclear magnetic resonance, infrared conductivity. The opening
of the gap indicates the decrease of the density of state around Fermi
level. We suppose that the particle gap results from the localization
of a partial of carriers. Below the crossover temperature $T^{*}$,
part of the carriers bind to the local Cu moments into localized Zhang-Rice
singlets. The carriers and the singlets coexist and transform into
each other with the same chemical potential. The transform process
bases on the Kondo-like interaction between carriers and the localized
Cu holes. 

Using the fermionic representation of the carrier's $\mathbf{s}$
and Cu's $\mathbf{S}$ operators, the Kondo-like term in dual fermion
model is rewritten as 
\begin{eqnarray}
H_{K}&=&J_{K}\sum_i\mathbf{s}_{i}\cdot\mathbf{S}_{i}\nonumber\\
&=&\frac{J_{K}}{4}\sum_{i\alpha\beta\gamma\delta}c_{i\alpha}^{\dagger}\sigma_{\alpha\beta}c_{i\beta}\cdot d_{i\gamma}^{\dagger}\sigma_{\gamma\delta}d_{i\delta}.
\end{eqnarray}

Since the energy level of the singlet is located about $J_K$ (at eV energy scale) below that of the triplet, we ignored the triplet channel at low temperature. Using the triplet and singlet decomposition of the Pauli matrix production in the Eq.~(\ref{PTS}), in the singlet channel, the Kondo-like interaction reads 
\begin{eqnarray}
H_{K}=-\frac{3J_{K}}{8}\sum_i\left(c_{i\uparrow}^{\dagger}d_{i\downarrow}^{\dagger}-c_{i\downarrow}^{\dagger}d_{i\uparrow}^{\dagger}\right)\left(d_{i\downarrow}c_{i\uparrow}-d_{i\uparrow}c_{i\downarrow}\right).
\end{eqnarray}

We further introduce a composite fermionic operator to describe the annihilation of a localized Zhang-Rice singlet accompanying a creation of a Cu d hole or vice versa 
\begin{eqnarray}
h_{i\sigma}=\frac{1}{\sqrt{2}}d_{i\sigma}^{\dagger}\left(d_{i\downarrow}c_{i\uparrow}-d_{i\uparrow}c_{i\downarrow}\right).
\end{eqnarray}
Thus, in the singlet channel, the Kondo-like interaction describes the transform process between the carriers and the composite fermions,
\begin{eqnarray}
H_{K}=-\frac{3\sqrt{2}J_{K}}{16}\sum_i\left[\left(c_{i\uparrow}^{\dagger}h_{i\downarrow}-c_{i\downarrow}^{\dagger}h_{i\uparrow}\right)+H.c.\right].
\end{eqnarray}

Clearly, the number of the composite fermions is equal to that of the localized Zhang-Rice singlets. The creation of a composite quasiparticle locally breaks its surrounding magnetic bonds and is equivalent to
a soliton on the spin-$1/2$ square lattice. Therefore, the composite quasiparticle forms a dispersionless energy level just at the Fermi level, the hybridization between the carriers and composite fermions induces an indirect gap as in Kondo-insulators \cite{Coleman2007} 
\begin{eqnarray}
\Delta_{ind}\approx2\left(\frac{3\sqrt{2}J_{K}}{16}\right)^{2}\frac{1}{D}\approx\frac{J_{K}^{2}}{7D}\sim J_{H}.
\end{eqnarray}

Considering the doping dependence of the $J_{H}$, $T^{*}\sim\Delta_{ind}\sim J_{H}(1-a\delta)$
(in ref. \cite{Barzykin2009}) with a constant $a$ and the hole doping
concentration $\delta$. The half-width of the conduction band $D$
is around 1 eV. 

Actually, in the strange metal phase, a temporal Zhang-Rice singlet
could form from an itinerant O hole and a localized d hole, and decomposes
into two holes repeatedly as the O hole hops from one site to another.
This by no means deviates away from our previous picture that the
carrier Fermi gas baths in the scattering of the localized Cu holes.

\setcounter{equation}{0}
\renewcommand{\theequation}{D\arabic{equation}}
\section{\label{appsc}Superconductivity Hamiltonian}

To understand the cuprate superconductivity, the key is to unveil
the interaction mediating the formation of Cooper pairs. As the doped
holes couple to the local moments and form localized Zhang-Rice singlets,
then the correlation between those doped holes and others are broken
off. However, an itinerant doped hole could not always completely
screen a local moment one-on-one. Thus, the doped holes could interact
with both localized and other itinerant holes.

In dual fermion model, as long as the carriers fail to completely screen the magnetic moments locally, the carriers on unit cell $i$ and $j$ could interact with each other by exchange of the Cu-spin fluctuation in terms of a four point vertex, written in real space as 
\begin{eqnarray}
\Gamma_{\alpha\beta,\gamma\delta}(i,j,\omega)=-\frac{J_{K}^{2}}{4}\chi_{d}(i,j,\omega)\sigma_{\alpha\beta}\sigma_{\gamma\delta}.\label{4vertex}
\end{eqnarray}

The interaction Hamiltonian is written in real space as 
\begin{eqnarray}
H_{sc}=J_{K}^{2}\sum_{ij}\chi_{d}(i,j,\omega)\mathrm{\mathbf{s}}_{i}\cdot\mathrm{\mathbf{s}}_{j}.
\end{eqnarray}
It is also denoted by the Feynman diagram in the SC region of the
phase diagram. When we only consider the nearest-neighbour coupling,
the interaction can be reduced to

\begin{eqnarray}
H_{sc}=J_{K}^{2}\chi_{d}(\left\langle i,j\right\rangle ,\omega)\sum_{\left\langle i,j\right\rangle }\mathrm{\mathbf{s}}_{i}\cdot\mathrm{\mathbf{s}}_{j},\label{eq:t-J}
\end{eqnarray}
where the nearest neighbour $\chi_{d}(\left\langle i,j\right\rangle ,\omega)$ is assumed to be space independent. For a local pair, the energy of a spin-triplet is about $J_{K}^{2}\chi_{d}(\left\langle i,j\right\rangle ,\omega)$ higher than that of a spin-singlet and thus the antiferromagnetic spin fluctuations favors spin-singlet pairing. Applying the triplet and singlet  decomposition in Eq.~(\ref{PTS}), in the singlet channel, the
Hamiltonian in Eq.~(\ref{eq:t-J}) is written as 
\begin{eqnarray}
&&H_{sc}=-\frac{3}{4}J_{K}^{2}\chi_{d}(\left\langle i,j\right\rangle ,\omega)\times\nonumber\\
&&\sum_{\left\langle i,j\right\rangle }\frac{1}{\sqrt{2}}\left(c_{j\downarrow}^{\dagger}c_{i\uparrow}^{\dagger}-c_{j\uparrow}^{\dagger}c_{i\downarrow}^{\dagger}\right)\frac{1}{\sqrt{2}}\left(c_{i\uparrow}c_{j\downarrow}-c_{i\downarrow}c_{j\uparrow}\right).
\end{eqnarray}

After transforming to momentum space, the interaction becomes 
\begin{eqnarray}
H_{sc}=\int\frac{d^{2}kd^{2}k'}{(2\pi)^{4}}J\left({\rm \mathbf{k}}-{\rm \mathbf{k}}'\right)c_{{\rm \mathbf{k}}\uparrow}^{\dagger}c_{-{\rm \mathbf{k}}\downarrow}^{\dagger}c_{-{\rm \mathbf{k}}'\downarrow}c_{{\rm \mathbf{k}}'\uparrow},\label{hint}
\end{eqnarray}
with 
\begin{eqnarray}
J\left({\rm \mathbf{k}}-{\rm \mathbf{k}}'\right)=-\frac{3}{2}J_{K}^{2}\chi_{d}(\left\langle i,j\right\rangle ,\omega) \times\nonumber\\ \left[\cos\left(k_{x}-k'_{x}\right)+\cos\left(k_{y}-k'_{y}\right)\right]
\end{eqnarray}
on a square lattice. The Cooper pairing potentials are symmetrized
with $J({\rm \mathbf{k}}-{\rm \mathbf{k}}')$ and $J({\rm \mathbf{k}}+{\rm \mathbf{k}}')$
in the singlet channel as \cite{Coleman2007}

\begin{eqnarray}
V_{{\rm \mathbf{k}},{\rm \mathbf{k}}'}=\frac{J\left({\rm \mathbf{k}}-{\rm \mathbf{k}}'\right)+J\left({\rm \mathbf{k}}+{\rm \mathbf{k}}'\right)}{2},
\end{eqnarray}
i.e. 
\begin{eqnarray}
V_{{\rm \mathbf{k}},{\rm \mathbf{k}}'}=&&-\frac{3}{2}J_{K}^{2}\chi_{d}(\left\langle i,j\right\rangle ,\omega)\nonumber\\ &&\times\left[\cos k_{x}\cos k'_{x}+\cos k_{y}\cos k'_{y}\right].
\end{eqnarray}

The pairing interaction can be further decoupled into $d$-wave and
$s$-wave components, 
\begin{eqnarray}
2\cos k_{x}\cos k'_{x}+2\cos k_{y}\cos k'_{y}=\gamma_{k}\gamma_{k'}+\gamma_{k}^{s}\gamma_{k'}^{s},
\end{eqnarray}
with the $d$-wave gap function$\gamma_{k}=\cos k_{x}-\cos k_{y}$,
and the extended $s$-wave gap function $\gamma_{k}^{s}=\cos k_{x}+\cos k_{y}$.

Thus, the pairing interaction in the $d_{x^{2}-y^{2}}$ channel is
\begin{eqnarray}
V_{{\rm \mathbf{k}},{\rm \mathbf{k}}'}^{d}=-\frac{3}{4}J_{K}^{2}\chi_{d}(\left\langle i,j\right\rangle ,\omega)\gamma_{k}\gamma_{k'}=-g\gamma_{k}\gamma_{k'}.
\end{eqnarray}

Various experimental studies as well as the spin-fluctuation theories
have demonstrated the $d$-wave symmetry pairing in cuprates. In dual
fermion model,  the $d$-wave symmetry pairing is favored due to the Cu-spin fluctuaiton pairing interaction. In the $d_{x^{2}-y^{2}}$ channel, the Eq.~(\ref{hint}) is rewritten
in a symmetric form as

\begin{eqnarray}
H_{dsc}=-g\left[\int\frac{d^{2}k}{4\pi^{2}}\gamma_{k}c_{k}^{\dagger}c_{-k}^{\dagger}\right]\left[\int\frac{d^{2}k'}{4\pi^{2}}\gamma_{k'}c_{k'}c_{-k'}\right].
\end{eqnarray}

Applying mean field method, a gap parameter is defined 
\begin{eqnarray}
\Delta_{sc}\equiv-g\int\frac{d^{2}k}{4\pi^{2}}\gamma_{k}\left\langle c_{k}^{\dagger}c_{-k}^{\dagger}\right\rangle .
\end{eqnarray}

Thus, a $d$-wave BCS mean field Hamiltonian is reached 
\begin{eqnarray}
H_{scm}=\Delta_{sc}\int\frac{d^{2}k}{4\pi^{2}}\gamma_{k}c_{k}c_{-k}+H.c.
\end{eqnarray}

Remarkably, as the four point vertex~(\ref{4vertex}) acts on the
same carrier in the normal state then the MFL-like self-energy is
obtained in the strange metal phase. Therefore, the strange metal
is caused by the same underlying physics that causes high-$T_{c}$
superconductivity.

\setcounter{equation}{0}
\renewcommand{\theequation}{E\arabic{equation}}
\section{\label{appgapequ}Revised BCS gap equation}

Although the normal state of cuprates is universally anomalous, the
superconductivity is quite `normal', i.e. BCS-like. Kogan further
demonstrated that the Homes scaling, as a matter of fact, indicates
another universal property of BCS superconductors including cuprates \cite{Homes2004,Kogan2013}. 

The BCS gap equation is written as \cite{Rickayzen1969}: 
\begin{eqnarray}
-\int\frac{d^{2}k'}{4\pi^{2}}\frac{V_{{\rm \mathbf{k}},{\rm \mathbf{k}}'}\Delta_{{\rm \mathbf{k}}'}}{\sqrt{\xi_{{\rm \mathbf{k}}'}^{2}+\Delta_{{\rm \mathbf{k}}'}^{2}}}\tanh\frac{\beta\sqrt{\xi_{{\rm \mathbf{k}}'}^{2}+\Delta_{{\rm \mathbf{k}}'}^{2}}}{2}=\Delta_{{\rm \mathbf{k}}} \label{egap}.
\end{eqnarray}
Clearly, for s-wave superconductivity, the pairing interaction is expected to be negative and nearly isotropic. However, the AF spin fluctuation is strongly momentum dependent, i.e. sharply peaked at or near the AF wave vector $\mathbf{Q}$. Quantitatively, according to equation \ref{egap}, one finds no solution for an $s$-wave gap parameter since the dominant interaction $V_{{\rm \mathbf{k}},{\rm \mathbf{k}}'}$ is positive at $\mathbf{k-k'\sim Q}$.

To further simplify the gap equation, we assume that the $d$-wave gap function $\gamma_{k}\approx2\gamma_{\varphi}=2\cos2\varphi$
with $\varphi=\arctan(k_{y}/k_{x})$. The gap equation becomes 
\begin{eqnarray}
g&&\int\frac{d\varphi}{2\pi}\int_{0}^{\hbar\omega_{0}}d\xi\frac{\rho_{c}(\xi)\gamma_{\varphi}^{2}}{\sqrt{\xi^{2}+\gamma_{\varphi}^{2}\Delta_{sc}^{2}}}\tanh\frac{\beta\sqrt{\xi^{2}+\gamma_{\varphi}^{2}\Delta_{sc}^{2}}}{2}\nonumber\\&&=1,
\end{eqnarray}
where the bare particle $\rho_{c}(\xi)$ is a constant value $\rho_{c}^{0}$
for an ideal two-dimensional Fermi gas with quadratic dispersion.
The $\hbar\omega_{0}$ is the integral cutoff energy.

In the limit $\Delta_{sc}(T\rightarrow T_{c})\rightarrow0$, the critical
temperature is given by
\begin{eqnarray}
k_{B}T_{c}\approx\hbar\omega_{0}e^{-\frac{1}{\lambda}},
\end{eqnarray}
with $\lambda=g\rho_{c}^{0}/2$ for the weak coupling $d$-wave superconductivity.

Different from the conventional BCS method, we show that the integral
cutoff energy should be 
\begin{eqnarray}
\hbar\omega_{0}=\min(E_{F},\hbar\omega_{c}),
\end{eqnarray}
because $0\leq\xi\leq E_{F}$ in gap equation at low temperature,
and the cutoff energy of non-zero $g$ is $\hbar\omega_{c}$. As $\hbar\omega_{c}$
is far less than the Fermi energy $E_{F}$, the integral upper limit
$\hbar\omega_{0}$ is equal to $\hbar\omega_{c}$. Thus, the BCS gap
equation is restored. However, in the superconductors with small superfluid density, for example, the lightly hole-doped cuprates, $\hbar\omega_{c}\geq E_{F}\approx2D\delta$,
the $T_{c}$ is given by 
\begin{eqnarray}
k_{B}T_{c}\approx2D\delta e^{-\frac{1}{\lambda}}.
\end{eqnarray}

Thus, the critical temperature is proportional to the doping level,
agreeing with the experimental Uemura law \cite{Uemura1989} and the
phase stiffness model \cite{Emery1995} as well as the recent work \cite{Valentinis2016}. Therefore, not only in the overdoped regime but also in underdoped regime, the revised BCS gap
equation is competent. However, it is worth noting that we have ignored
the Coulomb repulsion between pairing carriers. 


\end{document}